\documentclass[twocolumn,preprintnumbers,amsmath,superscriptaddress,amssymb,aps]{revtex4-1}

\usepackage{graphicx}
\usepackage{dcolumn}
\usepackage{bm}
\usepackage{amsmath}
\usepackage{amssymb}
\usepackage{graphicx}
\usepackage{indentfirst}
\usepackage{booktabs}
\usepackage{multirow}
\usepackage{colortbl}

\selectfont

\begin{document}
\title{Built-in electric field and strain tunable valley-related multiple topological phase transitions in VSiXN$_4$ (X= C, Si, Ge, Sn, Pb) monolayers}

\author{Ping Li}
\email{pli@xjtu.edu.cn}
\address{State Key Laboratory for Mechanical Behavior of Materials, Center for Spintronics and Quantum System, School of Materials Science and Engineering, Xi'an Jiaotong University, Xi'an, Shaanxi, 710049, China}
\address{State Key Laboratory for Surface Physics and Department of Physics, Fudan University, Shanghai, 200433, China}
\author{Xiao Yang}
\address{Faculty of Electronic Information Engineering, Huaiyin Institute of Technology, Huaian 223003, China}
\author{Qing-Song Jiang}
\address{Faculty of Electronic Information Engineering, Huaiyin Institute of Technology, Huaian 223003, China}
\author{Yin-Zhong Wu}
\email{yzwu@usts.edu.cn}
\address{School of Physical Science and Technology, Suzhou University of Science and Technology, Suzhou 215009, China }
\author{Wei Xun}
\email{xunwei@hyit.edu.cn}
\address{Faculty of Electronic Information Engineering, Huaiyin Institute of Technology, Huaian 223003, China}

\date{\today}

\begin{abstract}
The valley-related multiple topological phase transitions attracted significant attention due to their providing significant opportunities for fundamental research and practical applications. However, unfortunately, to date there is no real material that can realize valley-related multiple topological phase transitions. Here, through first-principles calculations and model analysis, we investigate the structural, magnetic, electronic, and topological properties of VSiXN$_4$ (X = C, Si, Ge, Sn, Pb) monolayers. VSiXN$_4$ monolayers are stable and intrinsically ferrovalley materials. Intriguingly, we found that the built-in electric field and strain can induce valley-related multiple topological phase transitions in the materials from valley semiconductor to valley-half-semimetal, to valley quantum anomalous Hall insulator, to valley-half-semimetal, and to valley semiconductor (or to valley-metal). The nature of topological phase transition is the built-in electric field and strain induce band inversion between the d$_{xy}$/d$_{x2-y2}$ and d$_{z2}$ orbitals at K and $\rm K'$ valleys. Our findings not only reveal the mechanism of multiple topological phase transitions, but also provides an ideal platform for the multi-field manipulating the spin, valley, and topological physics. It will open new perspectives for spintronic, valleytronic, and topological nanoelectronic applications based on these materials.
\end{abstract}

\maketitle
\section{Introduction}
Valley degree of freedom and related manipulations have become rising topics in recent years \cite{1,2}. It is coupled with the spin degree of freedom to exhibit extraordinary quantum effects \cite{Li1,3,4,5,Li2}, such as the valley spin Hall effect \cite{6,7}, and valley polarized quantum anomalous Hall effect (VQAHE) \cite{3,4,8,9,10}. The coupling effects are typically strong in the transition metal elements with localized d electrons, and the effects will be further enhanced with the decrease of dimensions. Therefore, two-dimensional (2D) transition metal materials provide a good opportunity to investigate the manifestation of quantum covariation effects of charge, spin, topological, and valley.

In the 2D hexagonal lattice material, the extreme value of the valence and conduction bands are appeared at the K and $\rm K'$ points, forming the valley degrees of freedom. Due to the effect of the inversion symmetry ($\hat{P}$) breaking together with spin-orbit coupling (SOC), the K and $\rm K'$ valleys have opposite spins but degenerate energies, which is called as spin-valley locking. As a result, the charged carriers in the two opposite valleys were selectively stimulated by the photons with left-hand and right-hand circular polarization. In addition, if the system further breaks time-reversal symmetry ($\hat{T}$), the valley degenerate will disappear at K and $\rm K'$ points. It is named as ferrovalley \cite{11}. Therefore, exploring ferrovalley materials is beneficial to promoting the development of spintronics and valleytronics.

Recently, a new 2D transition metal material MoSi$_2$N$_4$ family has been successfully synthesized in the experiment \cite{12}, and more than 60 stable ternary compounds with similar structures have been predicted theoretically \cite{13}. Such a family of materials has many interesting physical properties, including intrinsic magnetism, valley polarization, transport, and topology \cite{14,15,16,17,18}. It was noticed that the 2D Janus transition metal dichalcogenides (TMDs) were also experimentally synthesized \cite{19}. It is well known that the MoSi$_2$N$_4$ family is the combination of TMDs and SiN surface layers. Thence, Janus MoSi$_2$N$_4$ family is also very hopeful to be prepared. The two chalcogen layers are different, and the mirror symmetry in the Janus MoSi$_2$N$_4$ family is broken. The impact of spontaneous out-of-plane dipole (the built-in electric field) and strain on the covariation effect (charge, spin, topological, and valley) may be crucial, but has not been clearly understood yet.

In this work, we systematically investigate the built-in electric field and strain on the covariation effect of spin, topological, and valley freedom of the VSiXN$_4$ (X= C, Si, Ge, Sn, Pb) monolayers. It is found that the built-in electric field and strain effects can induce a series of topological phase transitions, such as valley semiconductor (VSC), valley-half-semimetal (VHSM), valley quantum anomalous Hall insulator (VQAHI) and valley-metal (VM). It essentially originated from the built-in electric field and strain-induced band inversion between the d$_{xy}$/d$_{x2-y2}$ and d$_{z2}$ orbitals at K and $\rm K'$ valleys. Our findings create pathways for valley-related multiple topological phase transitions and further microelectronic devices with the perfect valley, spin, and topology.

\section{STRUCTURES AND COMPUTATIONAL METHODS}
To explore the electronic and magnetic structures, we used the Vienna $Ab$ $initio$ Simulation Package (VASP) \cite{20,21,22} within the framework of the density functional theory (DFT) for the first-principles calculations. The exchange-correlation energy was described by the generalized gradient approximation (GGA) with the Perdew-Burke-Ernzerhof (PBE) functional \cite{23}. The plane-wave basis with a kinetic energy cutoff of 500 eV was employed, and $17\times 17\times 1$ and $25\times 25\times 1$ $\Gamma$-centered $k$ meshes were adopted for structural optimization and self-consistent calculations. A vacuum of 20 $\rm \AA$ was set along the c-axis, to avoid the interaction between the sheet and its periodic images. The total energy convergence criterion and the force were set to be 10$^{-7}$ eV and -0.001 eV/$\rm \AA$, respectively. To describe strongly correlated 3d electrons of V \cite{11,24}, the GGA+U method is applied. The Coulomb repulsion U is varied between 1 eV and 4 eV. To confirm the results, the calculations are also checked with Heyd-Scuseria-Ernzerhof (HSE06) method. To investigate the dynamical stability, the phonon spectra were calculated using a finite displacement approach as implemented in the PHONOPY code \cite{25}. The maximally localized Wannier functions (MLWFs) were employed to construct an effective tight-binding Hamiltonian to explore the Berry curvature, anomalous Hall conductivity (AHC), and edge states \cite{26}. The calculated the AHC, it performed the Berry curvature calculations using the formula
\begin{equation}
	\sigma_{xy} = C\frac{e^2}{h},
\end{equation}
\begin{equation}
	C= \frac{1}{2\pi} \int_{BZ} d^2k ~\Omega(\textbf{k}),
\end{equation}

\begin{equation}
	\Omega(\textbf{k})=-\sum_{n}f_{n}\sum_{n\prime \neq n}\frac{2Im \left \langle \psi_{nk} \mid v_{x} \mid \psi_{n\prime k} \right \rangle \left \langle \psi_{n\prime k} \mid v_{y} \mid \psi_{nk} \right \rangle}{(E_{n\prime}-E_{n})^2},
\end{equation}
where C is Chern number, $\Omega(\textbf{k})$ is the Berry curvature in the reciprocal space, $v_{x}$ and $v_{y}$ are operator components along the x and y directions, and $f_{n}=1$ for the occupied bands, respectively \cite{39,40,41}. One can obtain the Chern number as well as AHC by integrating the Berry curvature in the entire Brillouin zone (BZ). Therefore, the edge states were calculated in a half-infinite boundary condition using the iterative Green's function method by the package WANNIERTOOLS \cite{27}.

\section{RESULTS AND DISCUSSION }	
\subsection{Structure and stability}
The crystal structure of the VSiXN$_4$ monolayer is shown in Fig. 1(a). VSiXN$_4$ consists of septuple layers of N-Si-N-V-N-X-N, with atoms in each layer forming a 2D hexagonal lattice. Each V atom is coordinated with six N atoms, forming a trigonal prismatic configuration, and then this VN$_2$ layer is sandwiched by Si-N and X-N layers. The space group of VSiXN$_4$ is P3m1 (No. 156), and the space inversion symmetry is broken [except VSi$_2$N$_4$, it is P-6m2 (No. 187)]. The lattice constant of VSi$_2$N$_4$ is optimized to 2.88 $\rm \AA$, agreeing well with previous work \cite{12,14}. However, the lattice constant of VSiCN$_4$, VSiGeN$_4$, VSiSnN$_4$, and VSiPbN$_4$ are optimized to 2.75 $\rm \AA$, 2.95 $\rm \AA$, 3.04 $\rm \AA$ and 3.07 $\rm \AA$ (See Table I), respectively. The lattice constant is increasing with the increase of the X atomic order. To confirm the stability of VSiXN$_4$ monolayers, the phonon spectra are calculated. As shown in Fig. S1, the absence of imaginary frequencies confirms that the VSiXN$_4$ monolayers are dynamically stable. Only VSiPbN$_4$ has a small imaginary frequency at the $\Gamma$ point, it is the numerical error. In previous reports \cite{Ge,LuoWei}, it is a common problem for 2D materials. These imaginary frequencies become smaller as we increase the supercell in the phonon spectrum calculations. In addition, as shown in Fig. S2, we calculated the formation energy for 2H and 1T phase VSiXN$_4$. The formation energy is expressed as E$_f$ = (E$_{tot}$ - $\mu_V$ - $\mu_X$ - 4$\mu_N$)/N, where E$_{tot}$ is the total energy of the VSiXN$_4$. The $\mu_V$, $\mu_X$, and $\mu_N$ are the chemical potential V, X, and N atoms, respectively. N is the number of atoms in VSiXN$_4$. As listed in Table SI, the negative value, -6.390 eV $\thicksim$ 7.504 eV, indicates that the VSiXN$_4$ lattice is a strongly bonded network and hence favors its experimental synthesis. Even though 1T phase formation energy of VSiGeN$_4$, VSiSnN$_4$, and VSiPbN$_4$ is lower than that of the 2H phase formation, the difference is very small. In addition, to confirm the dynamical stability of the 1T phase VSiXN$_4$, we calculated the phonon spectrum, as shown in Fig. S3. We found that the phonon spectrum of 1T phase VSiXN$_4$ has a large imaginary frequency. It indicates that the dynamics of 1T phase VSiXN$_4$ are unstable.

\begin{figure}[htb]
\begin{center}
\includegraphics[angle=0,width=0.8\linewidth]{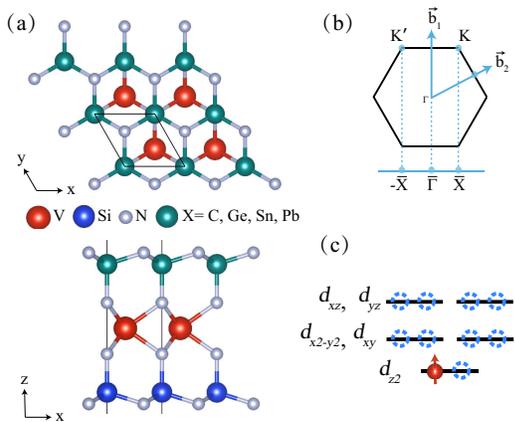}
\caption{(a) The top and side views of the crystal structure for VSiXN$_4$ (X = C, Si, Ge, Sn, Pb) monolayers. The red, gray, blue and green balls represent V, N, Si, and X elements, respectively. (b) The Brillouin zone (BZ) of the honeycomb lattice with the reciprocal lattice vectors $\vec{b}_1$ and $\vec{b}_2$. $\Gamma$, K, and M are the high-symmetry points in the BZ, and $\overline{\Gamma}$ and $\overline{X}$ are the high-symmetry points in the one-dimensional BZ. (c) The splitting of d orbitals under the trigonal prismatic crystal field. }
\end{center}
\end{figure}

\begin{table}[htbp]
\caption{
The calculated lattice constants a (\AA) for the monolayer, magnetic anisotropy energy (MAE) ($\mu$eV/cell), the valley degeneracy splits for the valence [$E_{v}^{K}$ - $E_{v}^{K'}$ (meV)], and conduction bands [$E_{c}^{K}$ - $E_{c}^{K'}$ (meV)], and global band gap E$_g$ (meV) of the 2D material VSiXN$_4$ (X = C, Si, Ge, Sn, Pb).}
\begin{tabular}{cccccccc}
	\hline
	               & a         & MAE        & $E_{v}^{K}$ - $E_{v}^{K'}$   & $E_{c}^{K}$ - $E_{c}^{K'}$        & E$_g$       \\
	\hline
	VSiCN$_4$      & 2.75      & 96.66      & 46.73                        &  0.00                             & 722.04        \\
	VSi$_2$N$_4$   & 2.88      & 50.98      & 63.74                        &  0.00                             & 381.33        \\
	VSiGeN$_4$     & 2.95      & -4.16      & 70.17                        &  0.00                             & 0.00        \\
	VSiSnN$_4$     & 3.04      & -17.41     &  0.00                        & 68.86                             & 118.91        \\
	VSiPbN$_4$     & 3.07      & -30.27     &  0.00                        & 72.45                             & 0.00         \\
	\hline
\end{tabular}
\end{table}	

\subsection{Magnetic property}
The valence electron configuration of the V atom is 3d$^3$4s$^2$. After donating four electrons to the neighboring N atoms, one valence electron is retained. According to Hund's rule and the Pauli exclusion principle, the electron configuration of V$^{4+}$ split into three groups: $a_1$ (d$_{z2}$ orbital), $e_1$ (d$_{xy}$, d$_{x2-y2}$ orbital), and $e_2$ (d$_{z2}$ orbital), as shown in Fig. 1(c). Therefore, the magnetic moment of the VSiXN$_4$ monolayer is expected to be 1 $\mu_B$ per cell. Our spin-polarized calculations indeed show that VSiCN$_4$, VSi$_2$N$_4$, VSiGeN$_4$, and VSiSnN$_4$ are spin polarized, and that the magnetic moments are mainly distributed over the V atoms, with a magnetic moment of 1 $\mu_B$ per unit. It is worth noting that only VSiPbN$_4$ the magnetic moment is not 1 $\mu_B$ (1.09 $\mu_B$). Due to the increase of the out-of-plane dipole formed by Si and Pb atoms, VSiPbN$_4$ becomes a VM.

\begin{figure}[htb]
\begin{center}
\includegraphics[angle=0,width=1.0\linewidth]{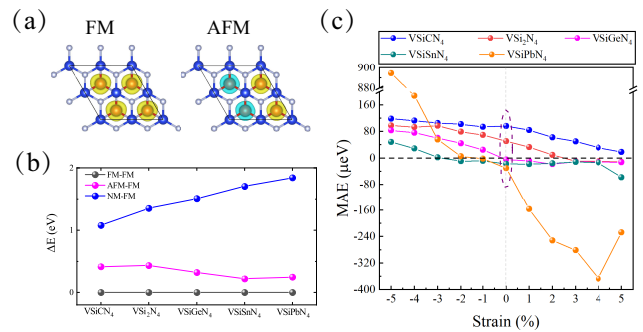}
\caption{(a) Spin charge densities with spin directions are indicated (yellow and cyan correspond to the spin up and spin down, respectively). The isovalue surface level is at 0.005 e/$\rm \AA$$^{3}$. (b) Calculated total energies of VSiXN$_4$ different magnetic structures, which is defined relative to that of FM state. (c) The magnetic anisotropy energy as function of strain. These results were obtained with U$_{\rm eff}$ = 3 eV.}
\end{center}
\end{figure}

To determine the magnetic ground state of the VSiXN$_4$ monolayer, three possible magnetic configurations are considered, namely, the ferromagnetic (FM), antiferromagnetic (AFM), and non-magnetic (NM) structures [see Fig. 2(a)]. We calculate the total energy difference between the FM, AFM, and NM using the GGA+U method. To determine a reasonable value of U, the GGA+U calculated energy difference between the AFM and FM states compares the HSE06 results, as shown in Fig. S4. It can be found that the value of U near 3 eV is consistent with the results of HSE06 for the VSiXN$_4$ monolayer. Therefore, we choose U$_{\rm eff}$ = 3 eV to investigate VSiXN$_4$ (X= C, Si, Ge, Sn, Pb) system. By comparing the total energy of FM, AFM, and NM states, we found that the FM configuration is the magnetic ground state for the VSiXN$_4$ monolayer, as shown in Fig. 2(b). The FM ground state of the VSiXN$_4$ monolayer can be understood by studying the crystal structure. In VSiCN$_4$, VSiGeN$_4$, VSiSnN$_4$, and VSiPbN$_4$, the V-N-V bond angles are 87.9$^\circ$, 91.4$^\circ$, 92.8$^\circ$ and 93.1$^\circ$, respectively, which are close to 90.0$^\circ$. According to the Goodenough-Kanamori-Anderson rule \cite{28,29,30}, this configuration is beneficial to FM coupling.

Then, we investigate the underlying physics for the robust out-of-plane magnetization. The direction of spin-polarization orientation is determined by magnetic anisotropy energy (MAE), which is defined as MAE = E$_z$ - E$_{x/y}$. Firstly, we tested the convergence of K-mesh before calculating MAE. As shown in Fig. S5, when the K-mesh is $25\times 25\times 1$, the convergence criterion is reached completely. The MAE is calculated using $25\times 25\times 1$ K-mesh in Fig. 2(c). The MAE value is listed in Table I. For magnetic ions V with a finite orbital moment, MAE can estimate through the formula $\lambda\langle L \rangle \langle S \rangle$, where $\lambda$, L, and S represent the strength of SOC, orbital angular momentum, and spin angular momentum \cite{31}, respectively. For the VSiXN$_4$ monolayer, the half-filled $a_1$ and empty e$_1$, e$_2$ orbitals indicate that L = 0 and S = 1/2. Due to the quenched orbital moment, the MAE originating from the SOC of V atom vanishes. Thence, MAE is mainly contributed by the SOC effect originating from N and X (X = C, Si, Ge, Sn, Pb) atoms. The MAE can be written as \cite{32,33,34}
\begin{equation}
\rm MAE=\xi^2 \sum_{u,o,\sigma,\sigma'}\sigma\sigma'\frac{|\langle o,\sigma|L_z|u,\sigma'\rangle|^2 - |\langle o,\sigma|L_x|u,\sigma'\rangle|^2}{E_{u,\sigma} - E_{o,\sigma'}},
\end{equation}
where o and u denote to the occupied and unoccupied states, respectively. The spin indices $\sigma$ and $\sigma'$ run over $\pm$1, amounting to each of the two orthogonal spin states at the K point. E$_{u,\sigma}$, and E$_{o,\sigma'}$ are the band energy of the states. As shown in Fig. S6, u and o are mainly contributed by N p$_x$ and p$_y$ orbitals. The contribution from the X (X = C, Si, Ge, Sn, Pb) atomic orbitals is almost negligible. Therefore, u and o can be written as a$_\tau$ = a$_x$ + i$\tau$a$_y$ from V atoms and p$_\tau'$ = p$_x$ + i$\tau'$p$_y$ from N atoms, where $\tau$ = $\pm$1 and $\tau'$ = $\pm$1. Thence, MAE can be simplified as
\begin{equation}
\rm MAE=\xi^2 \sum_{\tau, \tau'}\frac{|\langle a_\tau|L^N_z|p_{\tau'}\rangle|^2 - |\langle a_\tau|L^N_x|p_{\tau'}\rangle|^2}{\Delta},
\end{equation}
where $\Delta$ is the band gap, and L$_z|p_{\tau'}\rangle$ = $\tau'|p_{\tau'}\rangle$ and L$_x|p_{\tau'}\rangle$ = i$\tau'|p_{z}\rangle$,
\begin{equation}
\rm MAE=\xi^2 \sum_{\tau, \tau'}\frac{|\langle a_\tau|p_{\tau'}\rangle|^2 - |\langle a_\tau'|p_{z}\rangle|^2}{\Delta},
\end{equation}

The spin orientation is determined by the orbital overlaps of $\langle a_\tau|p_{\tau'}\rangle$ and $\langle a_\tau'|p_{z}\rangle$. In a purely octahedral crystal environment, the irreducible representations of the p$_\tau'$ and p$_z$ of the ligands are E$_u \oplus$E$_g$ and A$_{1g} \oplus$A$_{2u}$, respectively. a$_\tau$ of V atoms belongs to E$_g$, which indicates that a$_\tau$ trends to couple with p$_\tau'$ instead of p$_z$ from N atoms. Thus, $|\langle a_\tau|p_{\tau'}\rangle|$ $\gg$ $|\langle a_\tau'|p_{z}\rangle|$. It indicates that the MAE favors a positive value, benefiting the out-of-plane spin orientation. In fact, VSiCN$_4$ and VSi$_2$N$_4$ have an easy magnetization plane, while the easy axes of VSiGeN$_4$, VSiSnN$_4$, and VSiPbN$_4$ are along the out-of-plane direction, as shown in Fig. 2(c). It is attributed to the MAE tunable by the strong dipole interaction of the system. MAE is an intrinsic property of the material. There are many factors affecting the MAE of materials, such as spin-orbit coupling, structure, orbital occupation and so on. In VSiXN$_4$, MAE is mainly determined by the built-in electric field. As shown in Fig. 4(b), in the absence of strain, the VSiCN$_4$ has the largest built-in electric field. Besides, the built-in electric field is 0.36 V/$ \rm \AA$ (VSiCN$_4$), 0.16 V/$ \rm \AA$ (VSiGeN$_4$), 0.26 V/$\rm \AA$ (VSiSnN$_4$), and 0.31 V/$\rm \AA$ (VSiPbN$_4$), respectively. Exactly, the MAE is 96.66 $\mu$eV (VSiCN$_4$), -4.16 $\mu$eV (VSiGeN$_4$), -17.41 $\mu$eV (VSiSnN$_4$), and -30.27 $\mu$eV (VSiPbN$_4$), respectively. The variation law of MAE is completely consistent with that of the built-in electric field. In addition, as shown in Fig. 2(c), we found that the tensile strain favors an out-of-plane magnetism. It is originated from that the tensile strains reduce the distance between Si and X atoms, thus enhancing dipole interactions.

\subsection{Electronic band structure}
In addition, we first calculated the band structure of VSiXN$_4$ with spin polarization but without SOC. As shown in Fig. 3(a), it can be found that there is a valley at each of the K and $\rm K'$ points. Both the valence and conduction bands of the valleys are spin up bands. The two valleys are degenerate in energy. More interestingly, as the atomic number of X increases, it exhibits abundant electronic structure properties. VSiCN$_4$, VSi$_2$N$_4$, and VSiSnN$_4$ are VSC, while VSiGeN$_4$ and VSiPbN$_4$ are VHSM and VM, respectively. So we further calculated the band structures with the SOC effect. As shown in Fig. 3(b, c), it can be found that the valence band valley of K is appreciably higher than $\rm K'$ for the VSiCN$_4$, VSi$_2$N$_4$, and VSiGeN$_4$, while the conduction band valley of K is obviously lower than K$'$ for VSiSnN$_4$. Therefore, the valley degeneracy is broken and an evident valley splitting $E_{v}^{K}$ - $E_{v}^{K'}$ or $E_{c}^{K}$ - $E_{c}^{K'}$ is induced. $E_{v}^{K}$ - $E_{v}^{K'}$ (or $E_{c}^{K}$ - $E_{c}^{K'}$) is 46.73 meV, 63.74 meV, 70.17 meV, 68.86 meV and 72.45 meV for VSiCN$_4$, VSi$_2$N$_4$, VSiGeN$_4$, VSiSnN$_4$, and VSnPbN$_4$ (See Table I), respectively.

\begin{figure}[htb]
\begin{center}
\includegraphics[angle=0,width=1.0\linewidth]{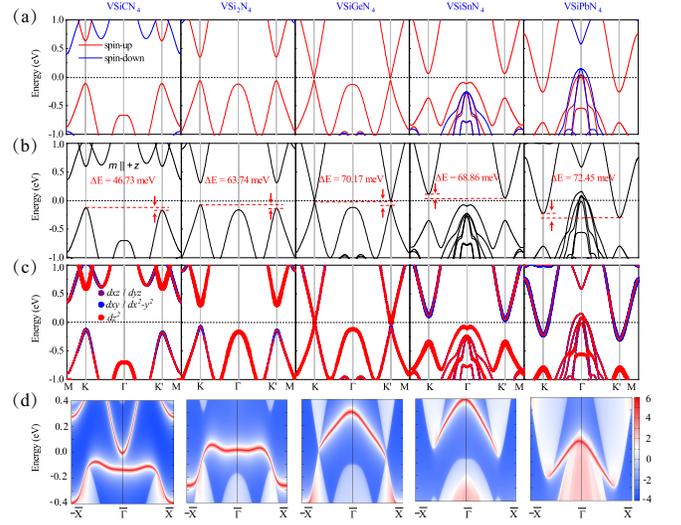}
\caption{ Band structures and edge state of VSiXN$_4$ monolayer obtained with the GGA+U (U$_{\rm eff}$ = 3 eV) method. (a) Spin polarized band structures of VSiXN$_4$ monolayer. The red and blue lines represent spin up and spin down bands, respectively. (b) Band structure with SOC of VSiXN$_4$ monolayer. (c) Orbital resolved band structure with SOC of VSiXN$_4$ monolayer. (d) The edge state of VSiXN$_4$ monolayer.}
\end{center}
\end{figure}

Here, it exhibits a novel phenomenon. The valley polarization of VSiCN$_4$, VSi$_2$N$_4$, and VSiGeN$_4$ only appears at the valence band, while it degenerates at the conduction band. Be the opposite, the valley splitting of VSiSnN$_4$ and VSiPbN$_4$ occurs at the conduction band. It is important that VSiGeN$_4$ is the critical state, and valence band maximum (VBM) and conduction band minimum (CBM) appear degeneracy at the K point forming a Dirac cone. From the orbital resolved band structure, as shown in Fig. 3(c), the VBM bands of VSiCN$_4$, VSi$_2$N$_4$, and VSiGeN$_4$ are mainly contributed by V d$_{xy}$/d$_{x2-y2}$ orbitals, while the CBM bands are dominated by d$_{z2}$ orbitals of V atom. For VSiSnN$_4$ and VSiPbN$_4$ monolayer, on the contrary, the VBM bands are primarily d$_{z2}$ orbital, while the CBM are mainly d$_{xy}$/d$_{x2-y2}$ orbitals. It can be found that the atomic number of the X atom induced band inversion between the d$_{xy}$/d$_{x2-y2}$ and d$_{z2}$ orbitals. As is well known, the orbital angular momentum of d$_{z2}$ orbital is zero. Thence, d$_{z2}$ orbital doesn't occur valley splitting. Notably, these results yield a good comparison with the band structures obtained from the HSE06 method, as shown in Fig. S7.

To understand the underlying mechanism for the ferrovalley effect in VSiXN$_2$. Here, we take VSiGeN$_4$ as an example to perform model analysis. We adopted $|$$\psi$$_v$$^{\tau}$$\rangle$=$\frac{1}{\sqrt{2}}$($|$d$_{xy}$$\rangle$+i$\tau$$|$d$_{x2-y2}$$\rangle$)$\otimes$$|$$\uparrow$$\rangle$, $|$$\psi$$_c$$^{\tau}$$\rangle$=($|$d$_{z2}$$\rangle$)$\otimes$$|$$\uparrow$$\rangle$ as the orbital basis for the VBM and CBM, where $\tau$ = $\pm$1 indicate the valley index corresponding to $\rm K/\rm K'$. Since the VBM and CBM belong to the same spin channel (spin up bands), we take the SOC effect as the perturbation term, which is
\begin{equation}
\hat{H}_{SOC} = \lambda \hat{S} \cdot \hat{L} = \hat{H}_{SOC}^{0} + \hat{H}_{SOC}^{1},
\end{equation}
where $\hat{S}$ and $\hat{L}$ are spin angular and orbital angular operators, respectively. $\hat{H}_{SOC}^{0}$ and $\hat{H}_{SOC}^{1}$ represent the interaction between the same spin states and between opposite spin states, respectively. For the VSiGeN$_4$ monolayer, the single valley is composed of only one spin channel [see Fig. 3(a)], and the other spin channel is far from the valleys. Hence, the term $\hat{H}_{SOC}^{1}$ can be ignored. On the other hand, $\hat{H}_{SOC}^{0}$ can be written in polar angles
\begin{equation}
\hat{H}_{SOC}^{0} = \lambda \hat{S}_{z'}(\hat{L}_zcos\theta + \frac{1}{2}\hat{L}_+e^{-i\phi}sin\theta + \frac{1}{2}\hat{L}_-e^{+i\phi}sin\theta),
\end{equation}
In the out-of-plane magnetization case, $\theta$ = $\phi$ = 0, then the $\hat{H}_{SOC}^{0}$ term can be simplified as
\begin{equation}
\hat{H}_{SOC}^{0} = \lambda \hat{S}_{z} \hat{L}_z,
\end{equation}
The energy levels of the valleys for the VBM and CBM can be expressed as E$_v$$^ \tau$ = $\langle$ $\psi$$_v$$^ \tau$ $|$ $\hat{H}$$_{SOC}^{0}$ $|$ $\psi$$_v$$^ \tau$ $\rangle$ and E$_c$$^ \tau$ = $\langle$ $\psi$$_c$$^ \tau$ $|$ $\hat{H}$$_{SOC}^{0}$ $|$ $\psi$$_c$$^ \tau$ $\rangle$, respectively. Then, the valley polarization in the valence and conduction bands can be expressed as
\begin{equation}
E_{v}^{K} - E_{v}^{K'} = i \langle d_{xy} | \hat{H}_{SOC}^{0} | d_{x2-y2} \rangle - i \langle d_{x2-y2} | \hat{H}_{SOC}^{0} | d_{xy} \rangle \approx 4\lambda,
\end{equation}
\begin{equation}
E_{c}^{K} - E_{c}^{K'} = 0,
\end{equation}
where the $\hat{L}_z|d_{xy} \rangle$ = -2i$\hbar$$|d_{x2-y2} \rangle$, $\hat{L}_z|d_{x2-y2} \rangle$ = 2i$\hbar$$|d_{xy} \rangle$. The analytical result certificates that the valley degeneracy splits for the valence and conduction bands are consistent with our DFT calculations ($E_{v}^{K}$ - $E_{v}^{K'}$ = 70.17 meV, $E_{c}^{K}$ - $E_{c}^{K'}$ = 0.00 meV).

\subsection{Built-in electric field induced topological phase transition}
To reveal the topological properties of VSiXN$_4$ monolayers, we have calculated the local density of states of the edge state through the Green's function method \cite{35}. As shown in Fig. 3(d), only the VSiGeN$_4$ monolayer exists a single topologically protected edge state appeared in between the conduction and valence bands. It indicates that only VSiGeN$_4$ could be topologically non-trivial, while the others are topologically trivial.

\begin{figure}[htb]
\begin{center}
\includegraphics[angle=0,width=1.0\linewidth]{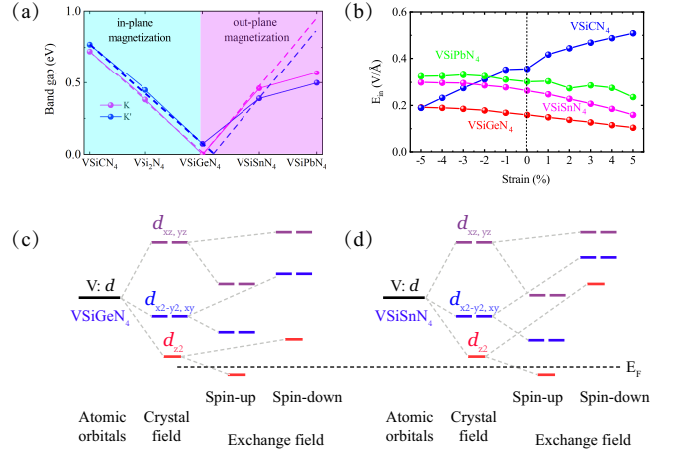}
\caption{(a) The band gap of VSiXN$_4$ (X = C, Si, Ge, Sn, Pb) at K and $\rm K'$ points. The dotted line is the fitting K and $\rm K'$ point gap variation trend. (b) The built-in electric field (E$_{in}$) as functions of strain for VSiXN$_4$ monolayer. [(c), (d)] The schematic illustration of energy level for V-d orbital in VSiGeN$_4$ (c) and VSiSnN$_4$ (d). E$_F$ represents the Fermi level. These results were obtained with U$_{\rm eff}$ = 3 eV.}
\end{center}
\end{figure}

It is noticed that an interesting topological phase transition from VSC to VHSM, to VSC, and to VM appears with the atomic number of X element increasing for the magnetic VSiXN$_4$ monolayers [see Fig. 3(b)] For instance, the topological phase transition occurs in VSiGeN$_4$ (VHSM). Through the detailed analysis on the energy band variation with the atomic number of the X element, it is found that the topological properties are closely related to the gap at the K and $\rm K'$ points. As shown in Fig. 4(a), as the atomic number of X increases, the gap at the K and $\rm K'$ points gradually decreases to zero in VSiGeN$_4$. The fitting curve shows that the band gap at the K and $\rm K'$ points would reach zero when X is Ge, which is the phase transition point. We further analyzed the built-in electric field. The built-in electric field is caused by the asymmetric Janus structure. Therefore, as shown in Fig. S8, the built-in electric field is defined as E$_{in}$ = ($\Phi_2$ - $\Phi_1$)/$\Delta h$, where $\Phi_1$ and $\Phi_2$ represent the electrostatic potential at the bottom and top of VSiXN$_4$, respectively. The $\Delta h$ is the structural height of VSiXN$_4$. As shown in Fig. 4(b), more surprisingly, the variation trend of the built-in electric field is completely consistent with the variation trend of the band gap of K and $\rm K'$. The built-in electric field is 0.36 V/$\rm \AA$ (VSiCN$_4$), 0.00 V/$\rm \AA$ (VSi$_2$N$_4$), 0.16 V/$\rm \AA$ (VSiGeN$_4$), 0.26 V/$\rm \AA$ (VSiSnN$_4$), and 0.31 V/$\rm \AA$ (VSiPbN$_4$), respectively. It means that the built-in electric field induces a topological phase transition. To prove this, we analyzed the energy level assignments of V-d orbitals for VSiGeN$_4$ and VSiSnN$_4$, as shown in Fig. 4(c, d). Seeing is believing. It is a band inversion between d$_{xy}$/d$_{x2-y2}$ and d$_{z2}$ orbitals at the K and $\rm K'$ points in VSiSnN$_4$. Note that the built-in electric field introduced by the element, which does not change continuously. VSiGeN$_4$ is the critical point, therefore, only VSiGeN$_4$ is topologically non-trivial. The continuous variation of the built-in electric field will be described in detail below. In addition, it is well known that applied electric fields can tune the band gap, magnetic ground state, topological properties, and so on. In previous reports \cite{Liu,Qian}, the electric field is an effective method to tune topological phase transition. The intrinsic built-in electric field tunes topological phase transition that is rarely reported.

\subsection{Strain induced topological phase transition}

\begin{figure}[htb]
\begin{center}
\includegraphics[angle=0,width=1.0\linewidth]{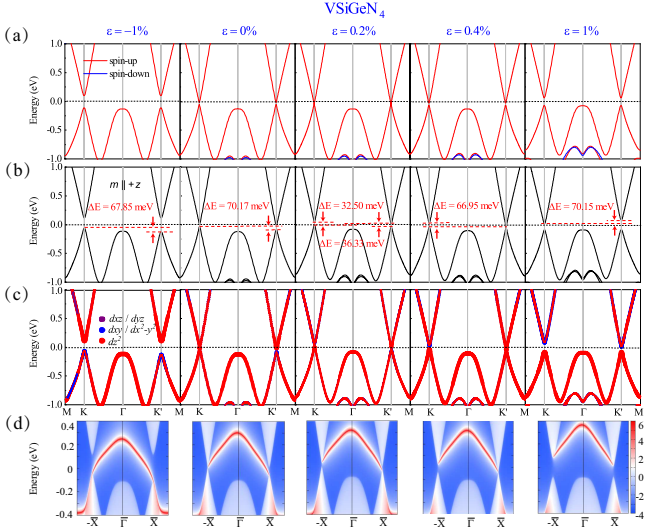}
\caption{ Band structures and edge state of VSiXN$_4$ monolayer obtained with the GGA+U (U$_{\rm eff}$ = 3 eV) method. (a) Spin polarized band structures of VSiGeN$_4$ monolayer with the biaxial strain. The red and blue lines represent spin up and spin down bands, respectively. (b) Band structure with SOC of VSiGeN$_4$ monolayer with the biaxial strain. (c) Orbital resolved band structure with SOC of VSiGeN$_4$ monolayer with the biaxial strain.(d) The edge state of VSiGeN$_4$ monolayer with the biaxial strain.}
\end{center}
\end{figure}

For a 2D material, its electronic structure can be generally tuned effectively by strain \cite{36,37,38}. In the following, we investigate the effect of biaxial strain on the spin, valley, and topological properties of VSiXN$_4$. Here, we focus on one representative VSiGeN$_4$. In the calculations, the biaxial strain is defined as $\varepsilon$ = (a-a$_0$)/a$_0$$\times$100$\%$. In the formula, a and a$_0$ represent lattice constant after and before in-plane biaxial strain is applied, respectively. As shown in Fig. 5(a-c), by increasing strain within a reasonable range (-5$\%$ $\thicksim$ 5$\%$), when $\varepsilon < 0 \%$, the material enters into a VSC. While $\varepsilon$ = 0 $\%$, the band gap at K point is firstly closed, meanwhile the band gap decreases to 70.17 meV at $\rm K'$ point. Hence, the VHSM states are acquired in the material. When employ the tensile strain, the band gap at K point reopens. At the other critical case with $\varepsilon$ = 0.4 $\%$, the band gap of $\rm K'$ point recloses, and the valleys at $\rm K'$ become a Dirac cone-shaped linear dispersion. Continuing applying the tensile strain to 0.4 $\%$, the band gap at the $\rm K'$ point reopens, and it again becomes a VSC.

\begin{figure}[htb]
\begin{center}
\includegraphics[angle=0,width=1.0\linewidth]{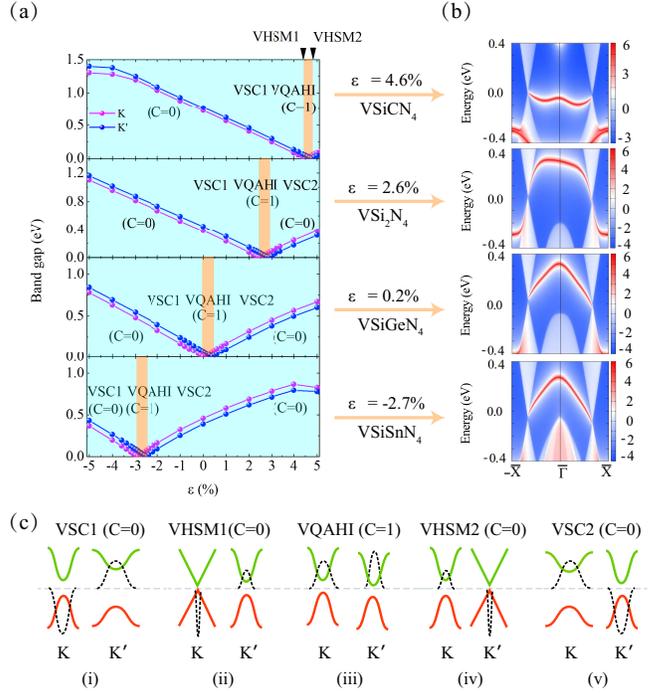}
\caption{ Band structures and edge state of VSiXN$_4$ monolayer obtained with the GGA+U (U$_{\rm eff}$ = 3 eV) method. (a) The band gap of VSiXN$_4$ at K and $\rm K'$ points with -5$\%$ $\thicksim$ 5$\%$ strain, and the orange and light blue shade denotes the VQAHI and VSC states, respectively. (b) The yellow shading corresponds to the edge states. (c) Schematic diagram of the evolution of the band structures and Berry curvatures of VSiXN$_4$ with the various strain. The green solid line, red solid line, and dotted lines represent the valence band, conduction band, and Berry curvatures of VSiXN$_4$ with the various strain, respectively. }
\end{center}
\end{figure}

To understand the mechanism of multiple topological phase transitions, we analyze the orbital projected band structures and berry curvature. As shown in Fig. 5(c) and Fig. S9, the compressive strain is 1$\%$, the VBM is dominated by the d$_{xy}$ and d$_{x2-y2}$ orbitals of V atoms, while the CBM mainly comes from the d$_{z2}$ orbital. The Berry curvatures at K and $\rm K'$ have opposite signs. More interestingly, the tensile strain is used (0.2 $\%$ tensile strain is shown in Fig. 5), and the band gap of the K point reopens. The band inversion occurs between d$_{xy}$/d$_{x2-y2}$ and d$_{z2}$ orbitals at the K point, while the orbital order doesn't change at the $\rm K'$ point. Surprisingly, the Berry curvature of K and $\rm K'$ become of the same sign. It is a characteristic of VQAHI. It is also confirmed by the calculation of edge states [See Fig. 5(d)]. After the band gap at $\rm K'$ is closed ($\varepsilon$ = 0.4 $\%$). Continuing applying the tensile strain, there is also a band inversion between d$_{xy}$/d$_{x2-y2}$ and d$_{z2}$ orbitals at the $\rm K'$ point. The signs of the Berry curvatures around the $\rm K'$ flip. Thus, the system returns to the VSC again.

\begin{figure}[htb]
\begin{center}
\includegraphics[angle=0,width=1.0\linewidth]{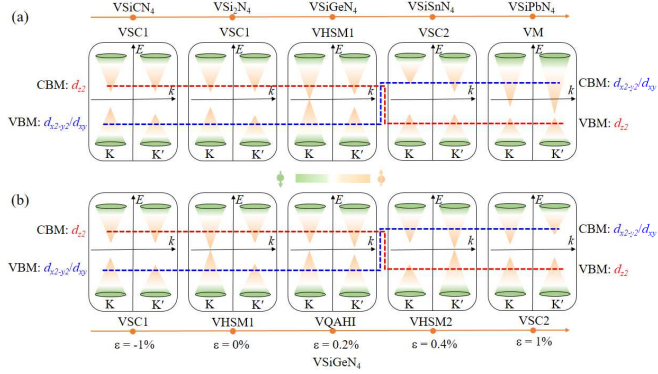}
\caption{Schematic diagrams of valley-dependent topological phase transitions. (a) Schematic diagram of the evolution of the band structures with the atomic number of X element increasing for VSiXN$_4$ monolayers. (b) Schematic diagram of the evolution of the band structures for the VSiGeN$_4$ monolayer as a function of strain. These results were obtained with U$_{\rm eff}$ = 3 eV.}
\end{center}
\end{figure}

To further demonstrate the universality of strain-induced band inversion mechanism in the VSiXN$_4$ system, we systematically investigated the topological properties of VSiXN$_4$ under different strains (band structure shown in Fig. S10-S14). As shown in Fig. 6(a,b), it is found that the strain can induce band inversions, which lead to the topological phase transition in these systems. Note that the range of the VQAHI states is different in VSiCN$_4$ (4.5$\%$ $\thicksim$ 4.8$\%$), VSi$_2$N$_4$ (2.5$\%$ $\thicksim$ 2.9$\%$), VSiGeN$4$ (0.0$\%$ $\thicksim$ 0.4$\%$), and VSiSnN$_4$ (-2.9$\%$ $\thicksim$ -2.5$\%$). To facilitate the reader's understanding, we also give a table, as listed in Table SII. Most important of all, the valley-related multiple topological phase transitions originate from the change of the sign of the Berry curvatures at K and $\rm K'$ points. How the band gaps and the topological phase vary with the various strain for the VSiXN$_4$ monolayers are summarized in Fig. 6(c). It is found that the strain magnitude required to achieve topological phase transition is highly dependent on the built-in electric field. We also found that the strain can effectively tune the built-in electric field, then, it induces the band inversion to realize topological phase transition. In addition, we calculated the results for different U values, as shown in Fig. S15. When the U$_{\rm eff}$ is 1 eV and 2 eV, the VSiXN$_4$ system doesn't have topological properties, while the U$_{\rm eff}$ increase to 3 eV and 4 eV, the VSiXN$_4$ system will appear topological phase with the built-in electric field and strain. Therefore, the band structures vary with the built-in electric field, and strain for VSiXN$_4$ monolayers is summarized in Fig. 7.

\begin{figure}[htb]
\begin{center}
\includegraphics[angle=0,width=1.0\linewidth]{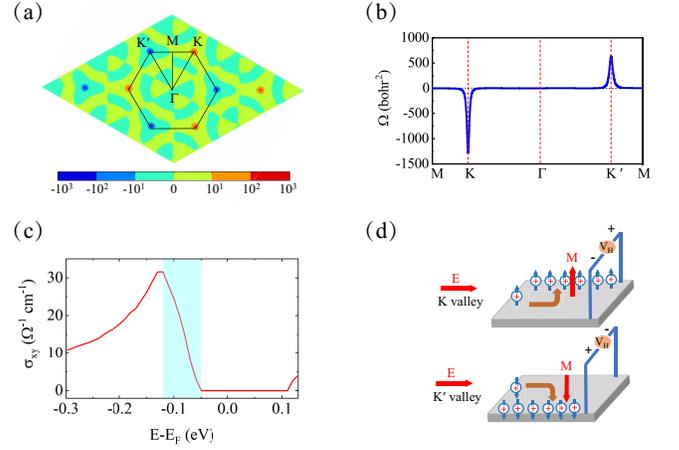}
\caption{(a) The Berry curvatures of VSiGeN$_4$ with the strain of -1$\%$ in the Brillouin zone and along the high symmetry line (b). (c) Calculated AHC $\sigma_{xy}$ as a function of Fermi energy. The light blue shadows denote the valley splitting between K and $\rm K'$ valley. (d) Schematic diagram of tunable the VQAHE in hole-doped VSiGeN$_4$ monolayer at the K and $\rm K'$ valley, respectively. The holes are denoted by the + symbol. Upward arrows and downward arrows refer to the spin up and spin down carriers, respectively. These results were obtained with U$_{\rm eff}$ = 3 eV.}
\end{center}
\end{figure}


To characterize the valley-contrasting physics in VSiXN$_4$ monolayer. We take the VSiGeN$_4$ of -1$\%$ strain as an example. The Berry curvature $\Omega(\textbf{k})$ of -1$\%$ strain VSiGeN$_4$ in the entire 2D BZ and along the high symmetry line are shown in Fig. 8(a, b). Clearly, the Berry curvatures at K and $\rm K'$ points have opposite signs, showing the typical valley polarization characteristic. By integrating the Berry curvature over the BZ, one can further calculate the AHC. As shown in Fig. 8(c), a valley-polarized Hall conductivity clearly exists in the -1$\%$ strain VSiGeN$_4$ monolayer. Specifically, when the Fermi level lies between the VBM or CBM of the K and $\rm K'$ valleys, as denoted by the cyan region, a fully spin- and valley-polarized Hall conductivity is generated. This result confirms the existence of valley anomalous Hall effect in the -1$\%$ strain VSiGeN$_4$ monolayer. Moreover, in the hole-doping condition, when the magnetism direction of VSiGeN$_4$ is in the +z direction, the spin up holes from the $\rm K'$ valley will be generated and accumulate on one boundary of the sample under an in-plane electrical field [upper plane of Fig. 8(d)]. On the other hand, when the magnetism direction is in the -z direction, the spin-up holes from the K valley will be generated and accumulate on the opposite boundary of the sample under an in-plane electrical field [lower plane of Fig. 8(d)]. This feature shows that monolayer VSiGeN$_4$ is an ideal candidate for the high-performance valleytronic devices.

\section{CONCLUSION}
In conclusion, we have demonstrated the rich multi-field induced physics in VSiXN$_4$ (X = C, Si, Ge, Sn, Pn) monolayers. The rich topological phase transitions can be realized through the built-in electric field and strain. Taking the VSiGeN$_4$ monolayer as an instance, when $\varepsilon <$ 0$\%$ and $\varepsilon >$ 0.4$\%$, it is a valley semiconductor. Moreover, the quantum anomalous Hall insulator is obtained with tensile 0.0 $\thicksim$ 0.4$\%$. At the two critical cases with $\varepsilon$ = 0.0$\%$ and 0.4$\%$, the valleys at K and $\rm K'$ become a Dirac cone, respectively. It becomes a valley-half-semimetal. We reveal that the nature of topological phase transition is built-in electric field and strain induces band inversion between the d$_{xy}$/d$_{x2-y2}$ and d$_{z2}$ orbitals at K and $\rm K'$ valleys. The abundant topological phase transition makes the VSiXN$_4$ monolayers a very promising material to develop intriguing spin-valley-topology devices.

\section*{ACKNOWLEDGEMENTS}
This work is supported by the National Natural Science Foundation of China (Grant No. 12004295). P. Li thanks China's Postdoctoral Science Foundation funded project (Grant No. 2022M722547), and the Open Project of State Key Laboratory of Surface Physics (No. KF2022$\_$09). This work was calculated at Supercomputer Center in Suzhou University of Science and Technology.


\end{document}